\documentclass[10pt,conference,letterpaper]{IEEEtran}

\usepackage{balance}  
\usepackage{graphicx} 
\usepackage{times}    
\usepackage{url}      
\usepackage{tabularx}
\usepackage{amsmath}
\usepackage{amssymb}
\usepackage{subfigure}
\usepackage{stfloats}
\usepackage{array}
\newcommand{\eat}[1]{}
\makeatletter
\def\url@leostyle{%
  \@ifundefined{selectfont}{\def\UrlFont{\sf}}{\def\UrlFont{\small\bf\ttfamily}}}
\makeatother
\def\pprw{8.5in}
\def\pprh{11in}

\eat{
\usepackage[pdftex]{hyperref}
\hypersetup{
pdftitle={SIGCHI Conference Proceedings Format},
pdfauthor={LaTeX},
pdfkeywords={SIGCHI, proceedings, archival format},
bookmarksnumbered,
pdfstartview={FitH},
colorlinks,
citecolor=black,
filecolor=black,
linkcolor=black,
urlcolor=black,
breaklinks=true,
}
}

\begin{document}

\title{Multi-scale Population and Mobility Estimation\\ with Geo-tagged Tweets}

\author{
Jiajun Liu, Kun Zhao, Saeed Khan, Mark Cameron, Raja Jurdak\\
{Commonwealth Scientific and Industrial Research Organisation (CSIRO), Australia}\\
{ \{firstname.lastname\}@csiro.au}\\
}

\maketitle

\begin{abstract}
Recent outbreaks of Ebola and Dengue viruses have again elevated the
significance of the capability to quickly predict disease spread in 
an emergent situation. However, existing approaches usually rely heavily on the
time-consuming census processes, or the privacy-sensitive call logs, leading to their
unresponsive nature when facing the abruptly changing dynamics in the
event of an outbreak. In this paper we study the feasibility of using large-scale
Twitter data as a proxy of human mobility to model and predict disease
spread. We report that for Australia, Twitter users' distribution
correlates well the census-based population distribution, 
and that the Twitter users' travel patterns
appear to loosely follow the gravity law at multiple scales of
geographic distances, i.e. national level, state level and
metropolitan level. The radiation model is also evaluated on
this dataset though it has shown inferior fitness as a result of
Australia's sparse population and large landmass. 
The outcomes of the study form the cornerstones for future work towards
a model-based, responsive prediction method from Twitter data for disease spread.
\end{abstract}

\section{Introduction}
Information on human mobility is essential for modeling and predicting disease spread.
This information can be obtained using traditional
census data \cite{census}, or new sensing technologies such as mobile phones \cite{mobile,cell}, RFID \cite{zhao2011social} and WIFI logging \cite{wifi}. 
For example, census data of commuting flows \cite{census} has been used in the simulation of multi-scale human mobility and spatial spread of infectious 
diseases. Traffic patterns in air transportation networks have been used to model human mobility in response to a disease
outbreak \cite{meloni2011modeling}. The study in \cite{cell}
reports that human mobility can be inferred by
the frequency of mobile phone calls between two locations and their
geographical distance. Mobile phone records coupled with malaria
prevalence data have been used to identify the dynamics of human
carriers of the parasite 
responsible for spreading disease across regions in
Kenya \cite{malaria}. Numerous methods of using mobile
phone records as a proxy for sensing human mobility are surveyed in \cite{mobile}.  More recently,
advances of sensory and information technologies have
also enabled us to study disease spread
at a fine-grained level by using RFID \cite{zhao2011social} or WIFI logging
\cite{wifi} to obtain high-resolution human contact patterns in a
limited range. However, due to privacy concerns for call records
and the tremendous human effort
involved for censuses, these data
sources are not always accessible, and they suffer from a range of limitations, such as the latency of census data, the low spatiotemporal resolution of mobile phone records, and the limited sensing range of RFID and WIFI logging. The acquisition of real-time, high spatiotemporal resolution and large-scale human mobility patterns for fine-grained modeling and predicting epidemic dynamics remains a challenge. 

Twitter appear to be a promising proxy to overcome this challenge. 
For example, geo-tagged tweets can provide high-precision location
information, and in the meanwhile offer comprehensive metadata with higher
granularity and resolution. Moreover, tweets are generated
continuously in large volume and are generally
available to public, which provides timely and accessible information on human mobility. Although twitter constitutes a valuable data source for
tracking human mobility, its potential power in modelling human mobility and epidemic spreading has not been fully exploited.

\vspace{-0.2cm}
\begin{figure}[thp]
\hspace{0.6cm}
\includegraphics[scale = 0.5]{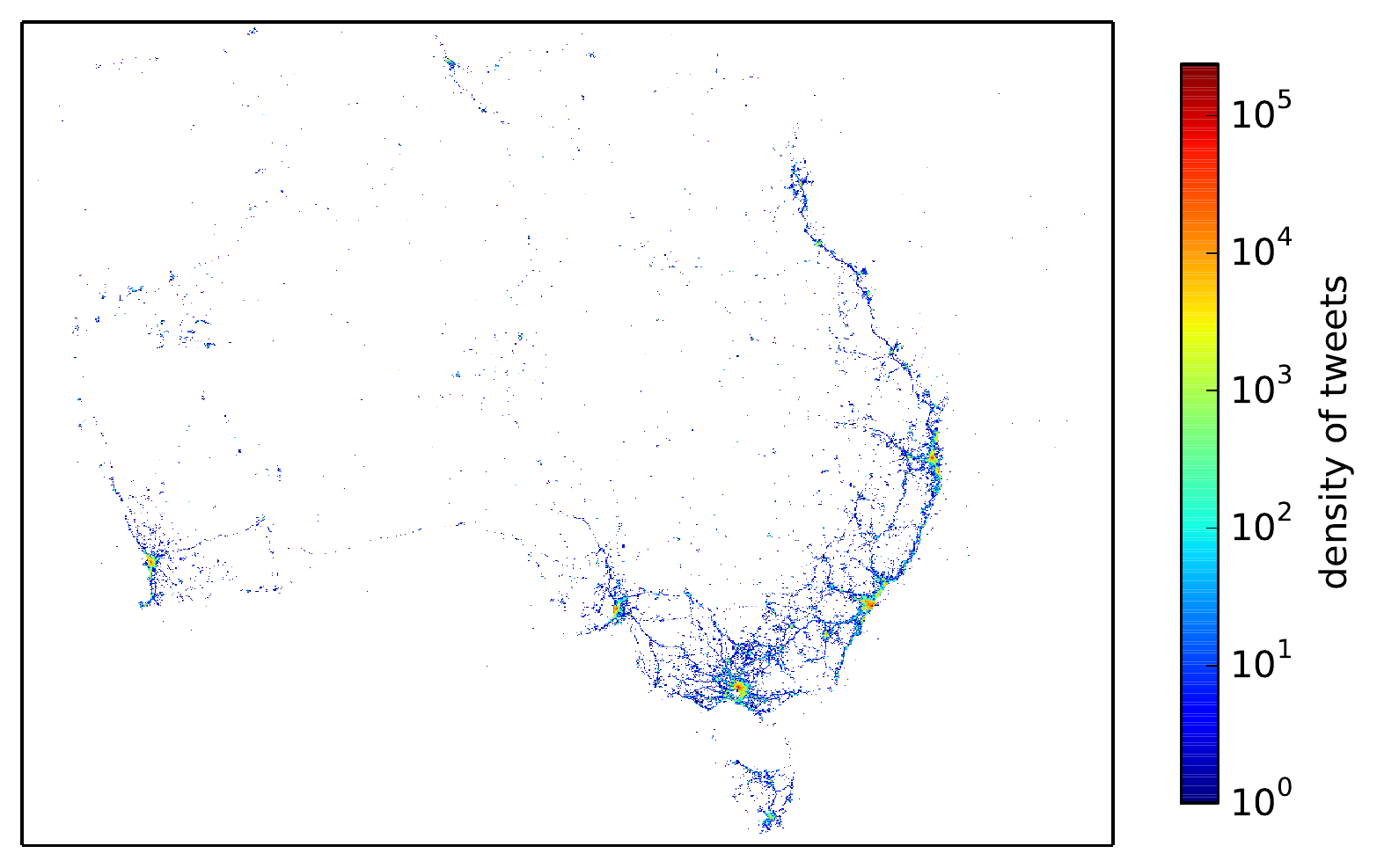}
\caption{The visualization of the geo-tagged Tweets
  highlights Australia's most dense areas and roughly resembles 
its population distribution.}
\label{fig:map}
\end{figure}
\vspace{-0.3cm}

In this paper, we report some preliminary results for the estimation 
of population distribution and human mobility flows in Australia using a large-scale dataset of 
6,304,176 geo-tagged Tweets generated by 473,956 unique users. 
We find that the population distribution can
be roughly estimated from geo-tagged Tweets. We apply two seminal models, namely gravity model and radiation model, to estimate mobility flows. In contrast to the
conclusion that the radiation model performs better than
the gravity model \cite{radiation}, the gravity model demonstrates superior
goodness of fit as well as better robustness to changes in geographic
scales in the estimation of mobility flows in Australia. The likely ingredients of such results include the unique characteristics of
Australia's geographic features and its sparse population distribution.

\section{Geo-tagged Tweets}
Though affected by the inevitable sampling bias as well as by the
individual tweeting dynamics, geo-tagged Tweets are generally
considered a high quality data source for capturing human mobility
\cite{twitter1,twitter2}. Particularly at the national or the global scale, the collective tweeting efforts are
believed to reveal the ``true'' mobility patterns to a great
extent \cite{twitter1}. In addition, geo-tagged Tweets exhibit a few favourable
properties that elude other data sources, such as its massive user
base, near-instantaneous updates, and public availability, all of
which are crucial for modeling an emergent 
disease outbreak.
\begin{table*}[htp]
\centering
\caption{Statistics of the Dataset}
\label{tbl:cmp}
\begin{tabular}{ |c|c|c|c|p{1.1cm}|p{1.2cm}|p{1.2cm}|p{1.8cm}| }
\hline  
\raisebox{-0.7\height}{Range of longitude} & \raisebox{-0.7\height}{Range of latitude} & \raisebox{-0.7\height} {Collection period} &
\raisebox{-0.7\height}{No.Tweets} & No.unique users & Avg.Tweets /user & Avg.waiting time &
Avg.no. locations/user \\
\hline
[112.921112,~159.278717] & [-54.640301,~-9.228820] & Sept.2013-Apr.2014
&6,304,176 & 473,956 & 13.3& 35.5hr& 4.76\\
\hline
\end{tabular}
\label{tbl:stats}
\end{table*}

\begin{figure}[thp]
    \subfigure[]{
    \includegraphics[scale = 0.21]{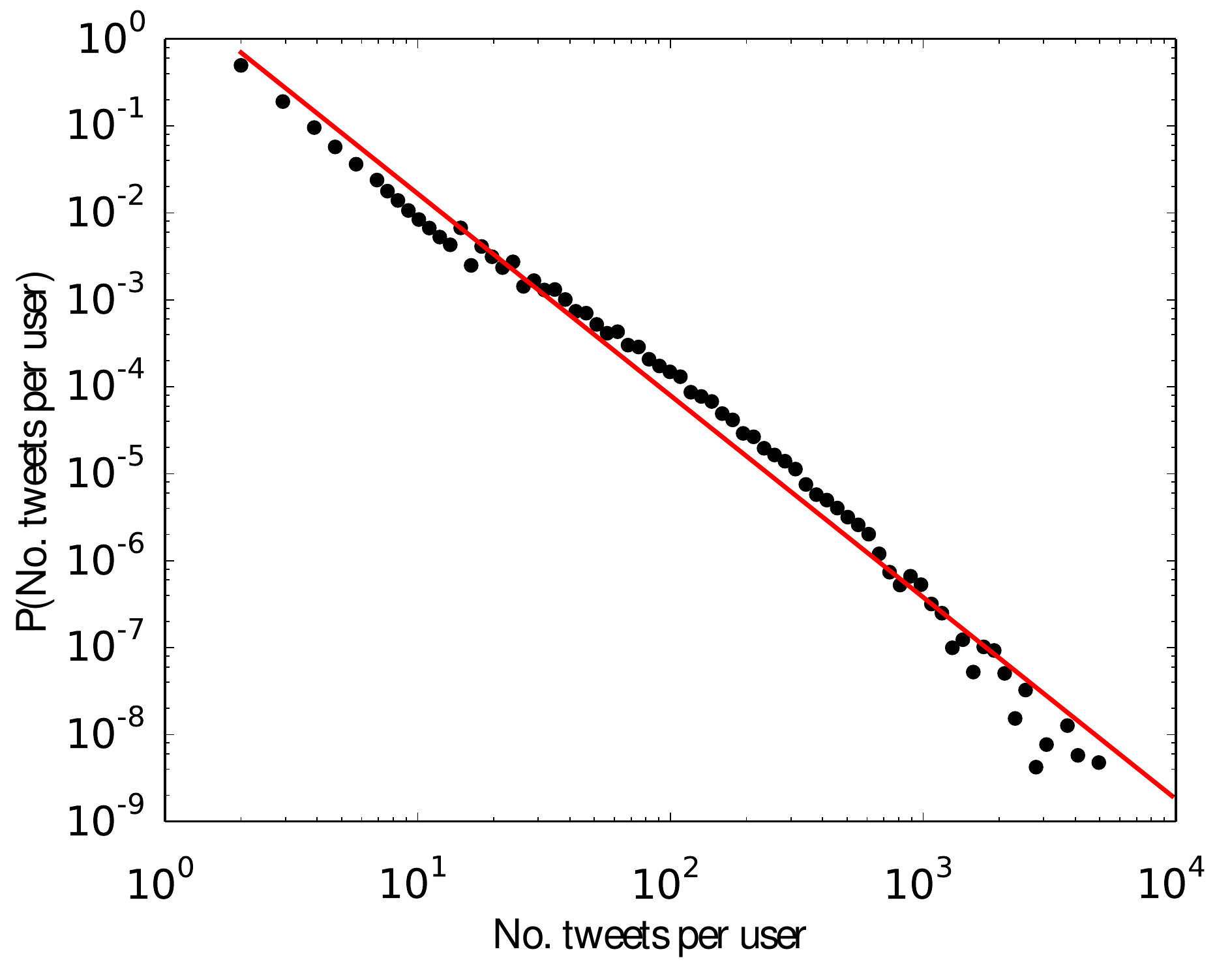}
  }
    \subfigure[]{
    \includegraphics[scale = 0.21]{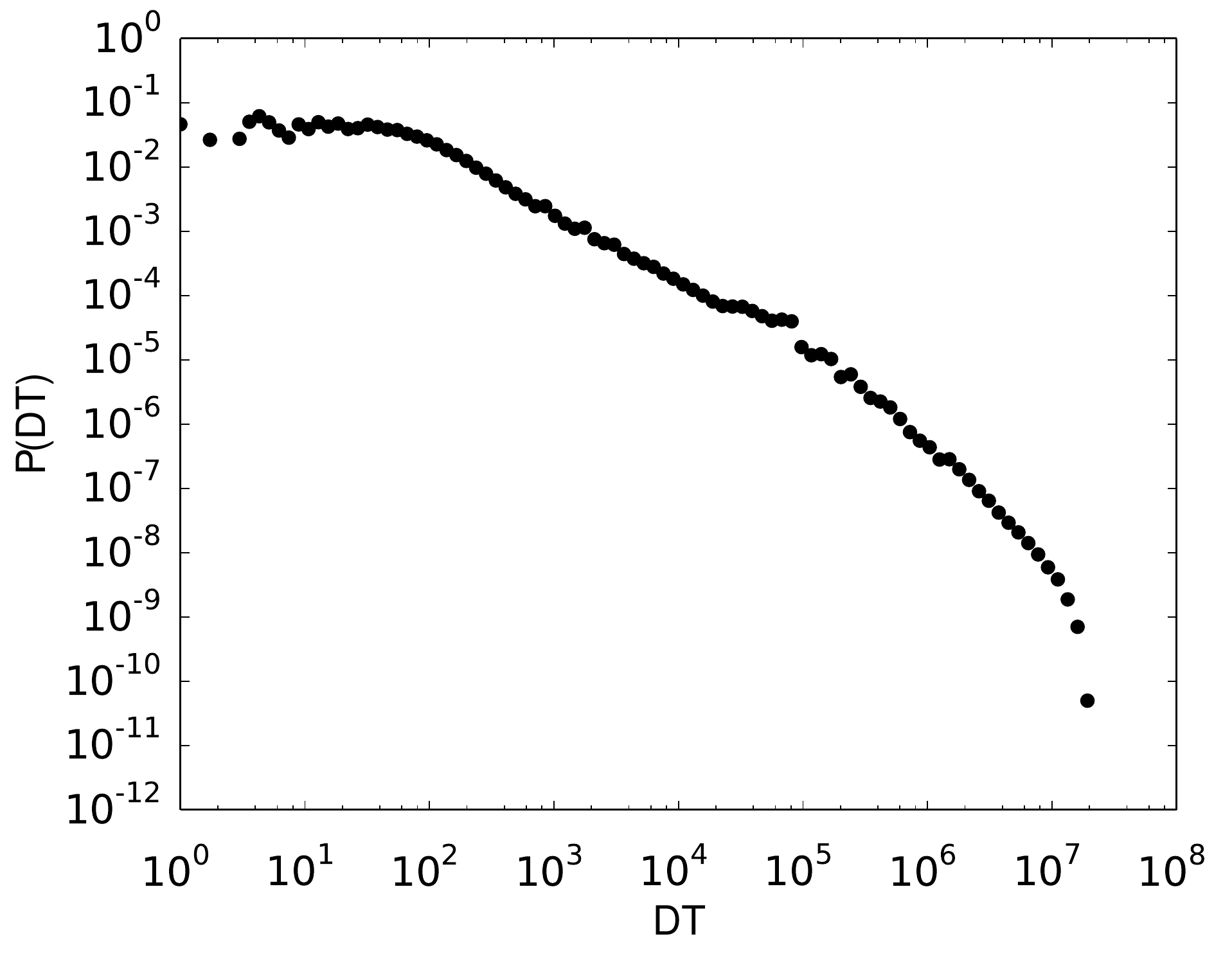}
  }

\caption{Both the number of Tweets per user and the time intervals
  between consecutive Tweets are heavy-tailed.}
\label{fig:nt}
\end{figure}

In light of these advantages, we collected all Tweets sent from
Australia from September 2013 to April 2014. We show the distribution of
the number of Tweets per user and the distribution of the waiting time
interval between consecutive Tweets to further explore the tweeting
dynamics in Figure \ref{fig:nt}. Both the distribution of the number of
Tweets per user and the distribution of waiting time span
at least eight decades and exhibit a heavy tail.  Evidently, similar to
other countries \cite{twitter1}, the tweeting behaviors of the Australian population
also exhibit the Pareto principle.  
 The distribution of the number of
Tweets per user essentially follows a power-law distribution
 whereas the distribution of time-intervals
demonstrates substantial heterogeneity which indicates that the
tweeting dynamics are likely to be influenced by multiple factors, such
as prioritized task handling in human behaviors \cite{bursts}.

\vspace{-0.5cm}
\hspace{-0.55cm}
\vspace{-0.5cm}

In Table \ref{tbl:stats} we provide more 
basic statistics of the collected Tweets. We use the longitude and
latitude ranges to filter the Tweets of interest published
from Australia. Over the 7 months collection
period, the average number of Tweets per user is only
13.3, however a small portion of the Twitter
population demonstrates enthusiasm in sharing their daily
life, with the numbers of users with more than 50, 100, 500, 1000 Tweets
being 23462, 10031, 766 and 180 respectively.

\section{Population Estimation}
To gain more insight into the feasibility of using Twitter data to
model and predict the spread of infectious diseases, 
we examine the correlation between the geographic distribution of Twitter
users and the Australian population distribution reported by census
\footnote{\url{http://www.abs.gov.au/AUSSTATS/abs@.nsf/DetailsPage/3218.02012-13}}. The
study is conducted at three scales, namely national level (20 most populated
cities in Australia), state level (20 most populated cities in the
state of New South Wales), and metropolitan level (20 most populated
suburbs in Sydney) to verify the robustness to geographic
scales. A search radius of 50km, 25km and 2km is used for the three
scales respectively when extracting number of Tweets, number of users and mobility
from Tweets. For the three scales, the average distances between
areas are 1422km, 341km and 7.5 km respectively.

\vspace{-0.3cm}
\begin{figure}[htp]
\hspace{-0.55cm}
   \centering
    \subfigure[$\epsilon=2km$]{
    \includegraphics[scale = 0.28]{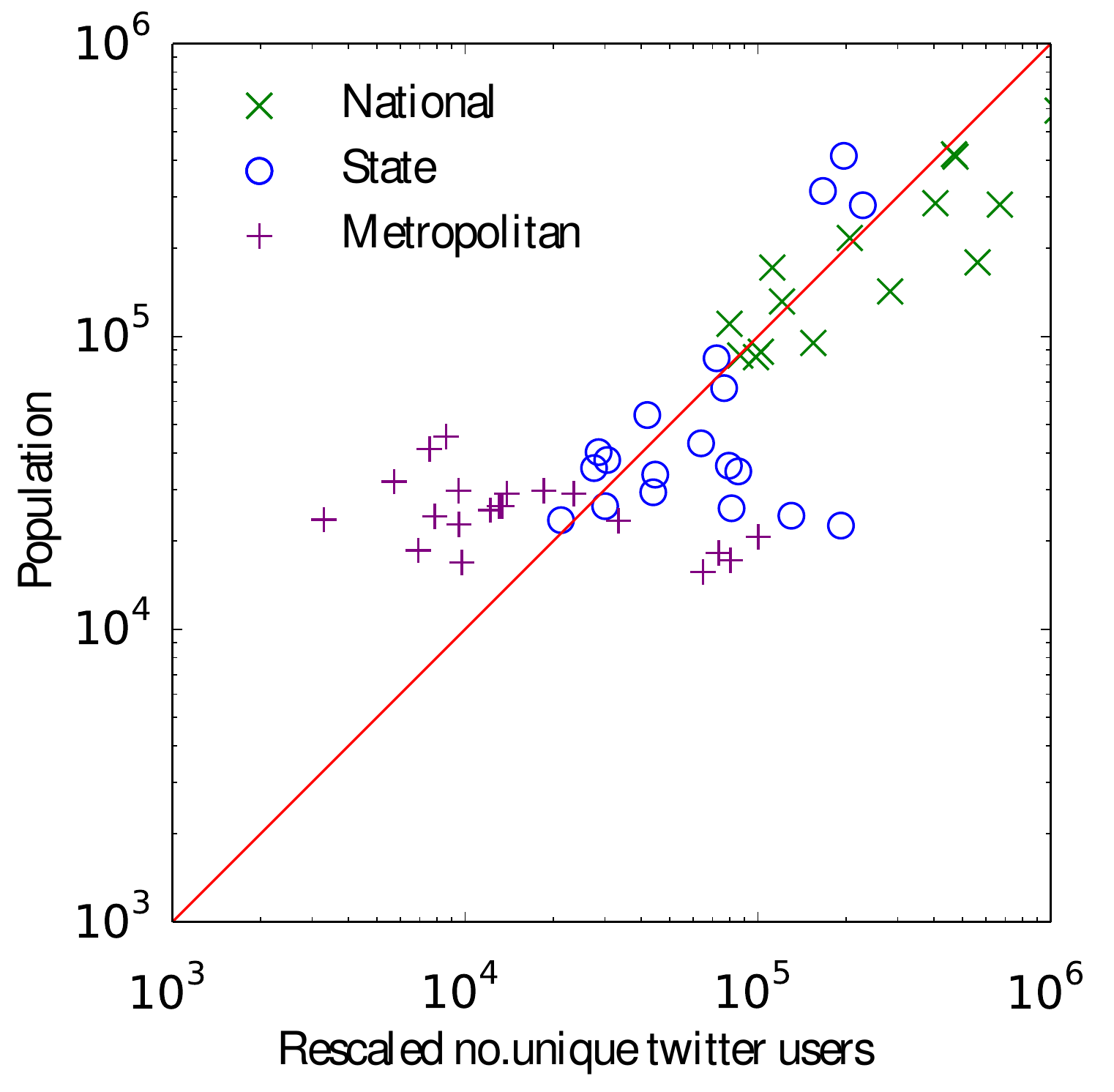}
    \label{fig:popua}
  }
    \subfigure[$\epsilon=0.5km$]{
    \includegraphics[scale = 0.28]{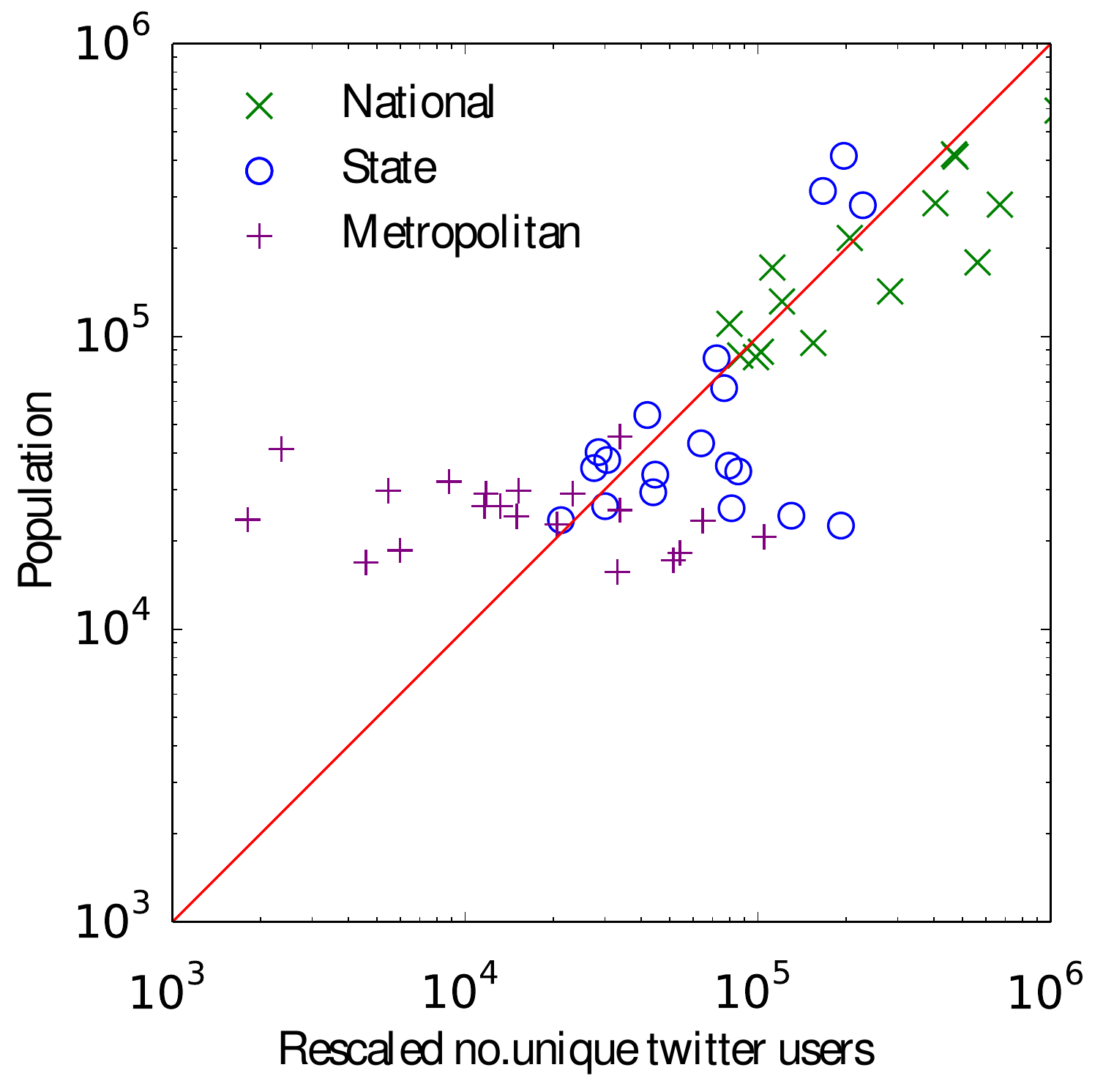}
    \label{fig:popub}
  }
\caption{The rescaled ``Twitter population'' ($Cp^{Twitter}\approx
  p^{Census}$, where $C$ is a rescaling factor) of the examined 60 areas
  at three different geographic scales shows strong correlation to
  the population distributions in the real-world. $\epsilon$ is the
  search radius to derive Twitter-based population.}
\end{figure}

Figure \ref{fig:popua} depicts the correlation between the number of unique Twitter
users and the census-based population of the examined areas at the three
scales. 
In total, 60 samples are shown in the figure,
with 20 points for each geographic scale. Overall we have a Pearson
correlation coefficient of 0.816 between the ``Twitter population
distribution'' and the real-world population distribution, with a
two-tailed p-value of $2.06\times 10^{-15}$, indicating a strong
correlation between the two. We argue that based on this result,
estimating the population distribution 
from the geo-tagged Tweets is feasible.

In addition, the correlation appears to
weaken as the population size and geographic scale decrease. That is,
the estimation for National aligns best with the census
data, and for State a few outliers begin to manifest, while for
Metropolitan the points scatter further more (though still
centered at $y=x$). It could be that the sample size of Tweets 
tend to decrease for smaller geographic areas, making the sample bias
a more significant factor. However, as we extract the median
numbers of users for the three scales, we find that for National,
State and Metropolitan, the figures are 4166, 743, and 3988
respectively. We observe that despite a smaller sample size, 
State yields better prediction power than Metropolitan, which
indicates that the sample size has limited influence on
the correlation. Now considering that the noises introduced while
conducting censuses and while extracting the Tweets and users 
tend to have greater effect on smaller areas, in attempt to investigate
such an effect, we change the search radius of Metropolitan to 
0.5km and generate Figure \ref{fig:popub}. This results in significant
increase of error for Metropolitan and hence we argue that the
sensitivity to the edges of the areas and search radius is 
likely to be a prominent factor to the differences of the correlation 
for different scales. A more in-depth discussion around this phenomenon
will be provided in future work.

\section{Mobility Estimation}
To explore the feasibility of 
mobility estimation with Twitter data, first we extract the mobility
from Tweets by counting how many pairs of consecutive Tweets appear
first at the source area and then the destination area, and then we
capture the mobility between areas using two models, namely the Gravity
model and the Radiation model:

\textbf{Gravity}: Similar to Newton's law of gravitation, the Gravity model suggests
that mobility between two places, namely an origin and a destination
is proportional to the product of populations of these two places, and
is inversely proportional to the power law of distance between them \cite{Zipf}. 
It is considered one of the fundamental models to predict not just
human mobility, but also trade flows between 
countries as well as communications volume between cities. 
However, it has also demonstrated some limitations. It is considered to work
well for transitions over shorter distances, but for larger 
distances the results are sometimes inconsistent, 
thus putting its universality in question.

\textbf{Radiation}: The Radiation model follows the analogy of the particle diffusion
model in which particles emitted at a source have certain probability
of being absorbed by a destination location. According to this model \cite{radiation}, 
the absorption probability depends on the origin's population and the destination's population. In addition, it is also determined by
the population within a circle centered at the origin, with the radius equal to the distance 
between the origin and the destination (excluding the origin and the
destination themselves). 

In our scenario, they are formally defined as:

\paragraph{Gravity Model (4 Parameters)}
\begin{equation}
P\propto C\frac{m^\alpha n^\beta}{d^\gamma}
\end{equation}
\paragraph{Gravity Model (2 Parameters)}
\begin{equation}
P\propto C\frac{m n}{d^\gamma}
\end{equation}
\paragraph{Radiation Model}
\begin{equation}
P\propto C\frac{mn}{(m+s)(m+n+s)}
\end{equation}
where $\alpha$, $\beta$, and $\gamma$ are the models' fitting
parameters, $C$ is the scaling parameter. $m$ and $n$ represent 
the population of the source and the
destination respectively. $d$ is the distance between the two areas. $s$ is defined as the total population within
radius $d$ from the center of the source area with the source and
destination areas excluded. For Gravity models, given
a series of $m$, $n$ and $d$ values, the parameters $\alpha$, $\beta$, and $\gamma$ can be estimated from
least-square fitting after taking logarithm of the formulas. In the
rest of the paper we refer to the three models as Gravity 4Param,
Gravity 2Param, and Radiation.

\vspace{-0.2cm}
\begin{figure}[htp]
\centering
\hspace*{-0.3cm}
    \subfigure[National]{
    \includegraphics[scale = 0.2]{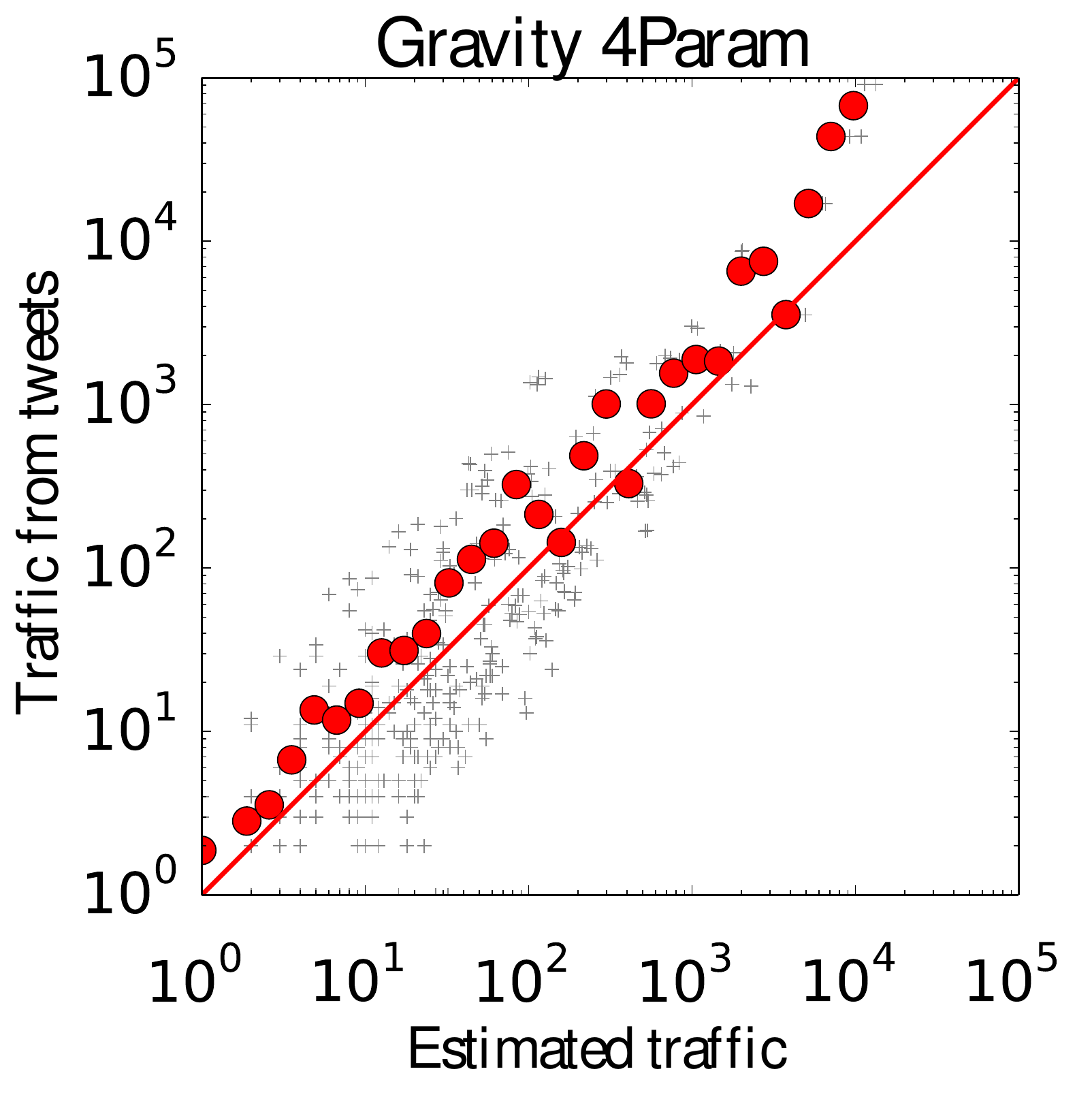}
    \includegraphics[scale = 0.2]{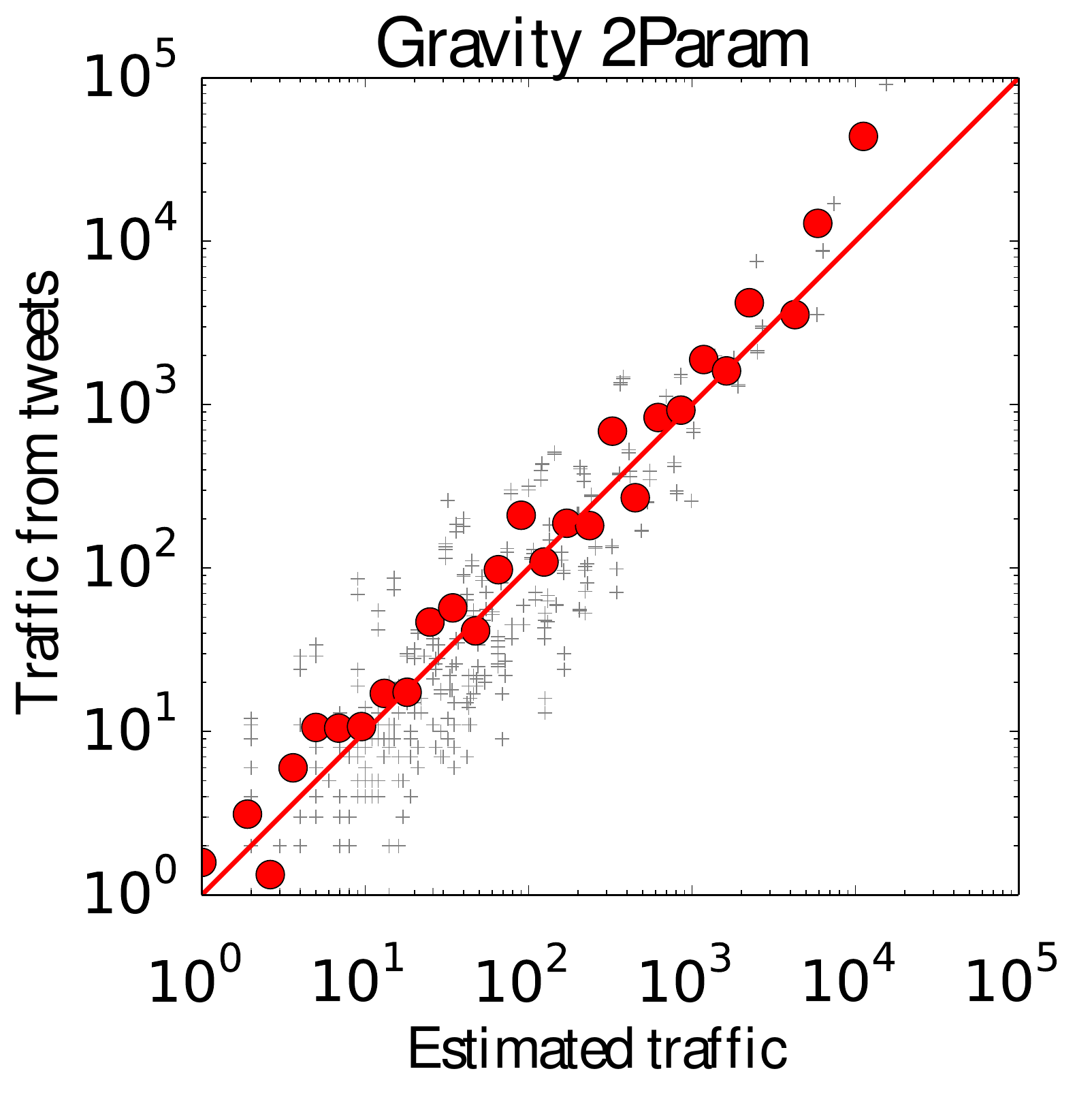}
    \includegraphics[scale = 0.2]{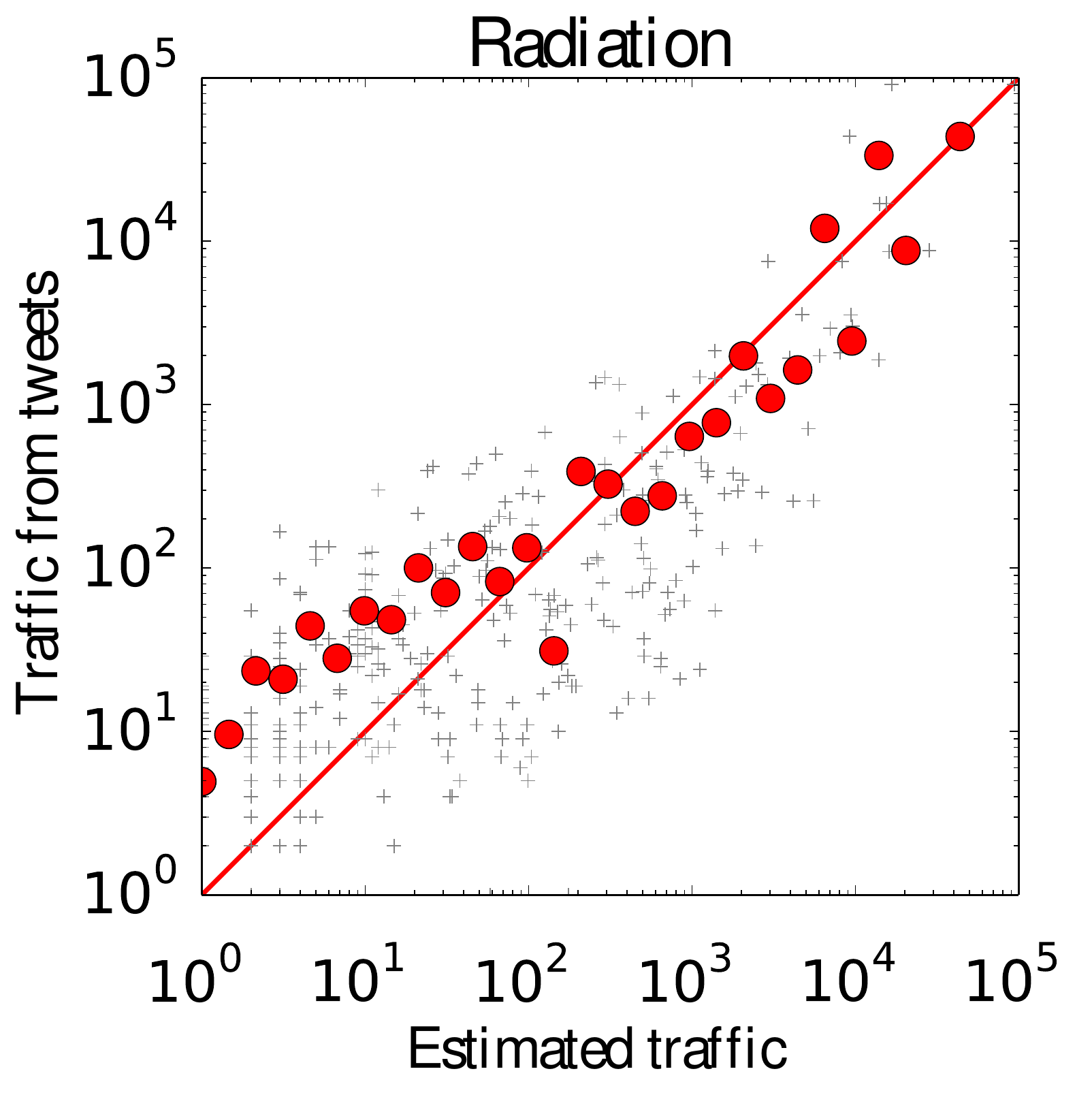}
    \label{fig:modela}
  }\\
\hspace*{-0.3cm}
    \subfigure[State]{
    \includegraphics[scale = 0.2]{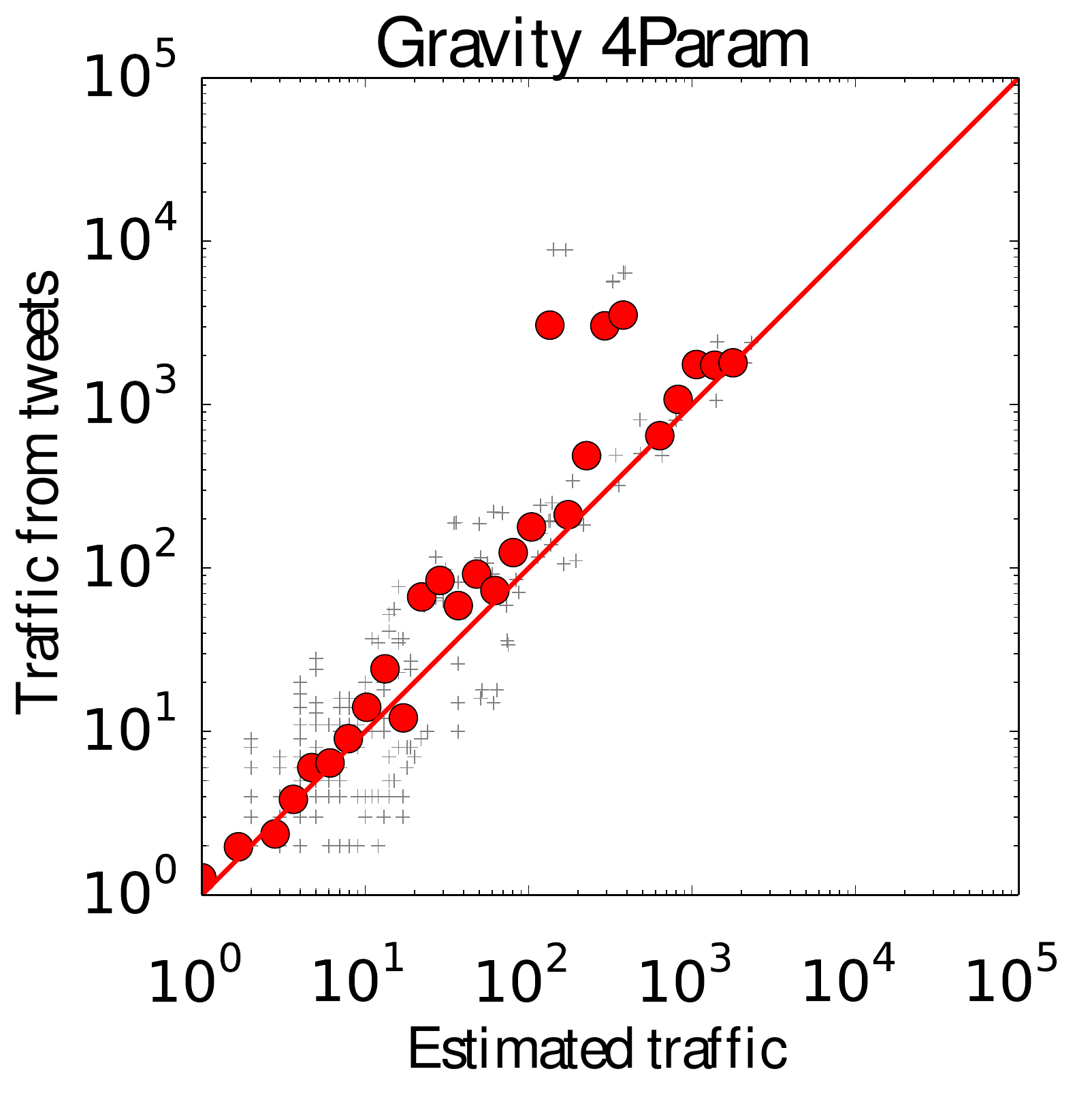}
    \includegraphics[scale = 0.2]{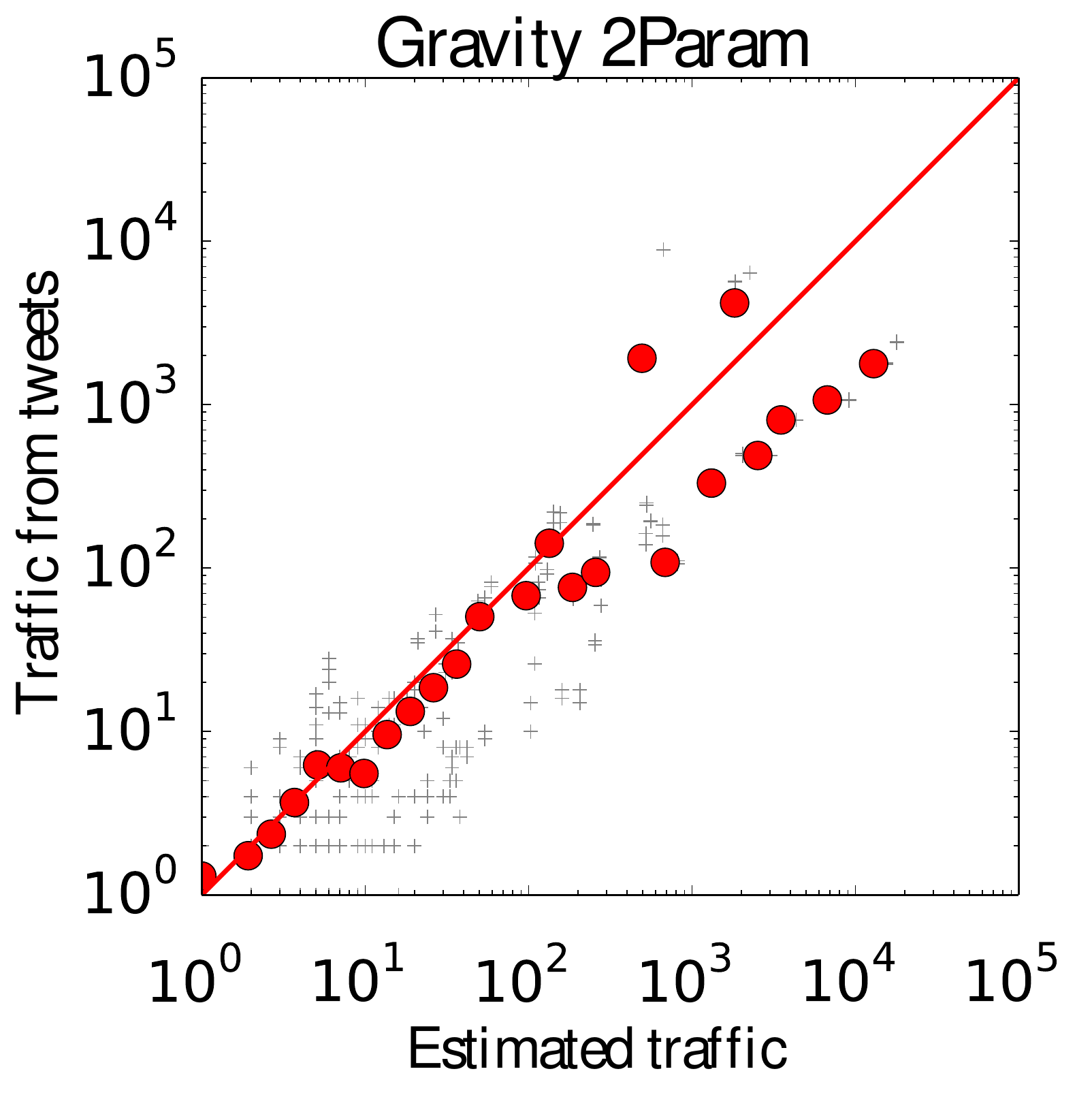}
    \includegraphics[scale = 0.2]{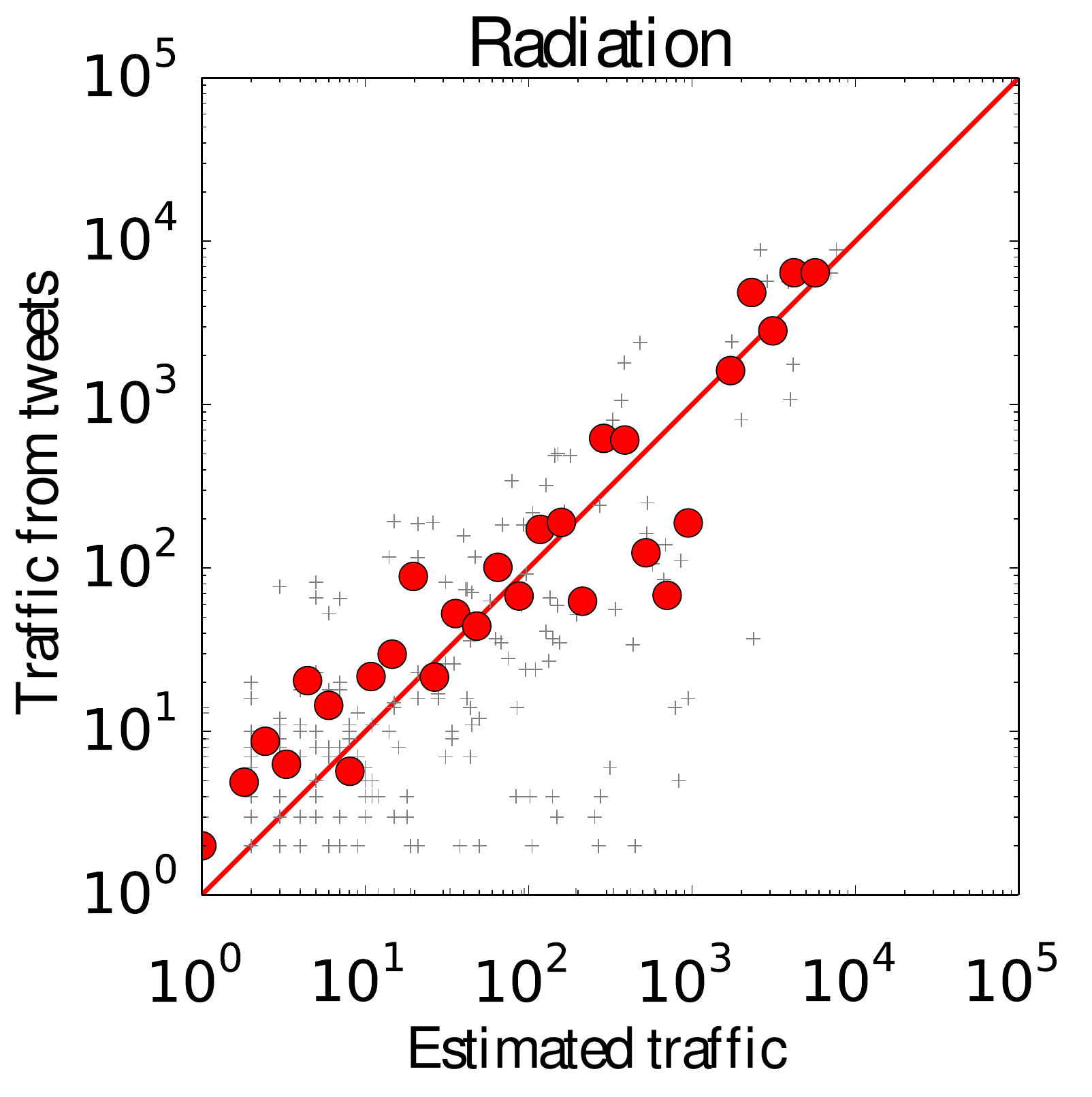}
    \label{fig:modelb}
  }\\
\hspace*{-0.3cm}
    \subfigure[Metropolitan]{
    \includegraphics[scale = 0.2]{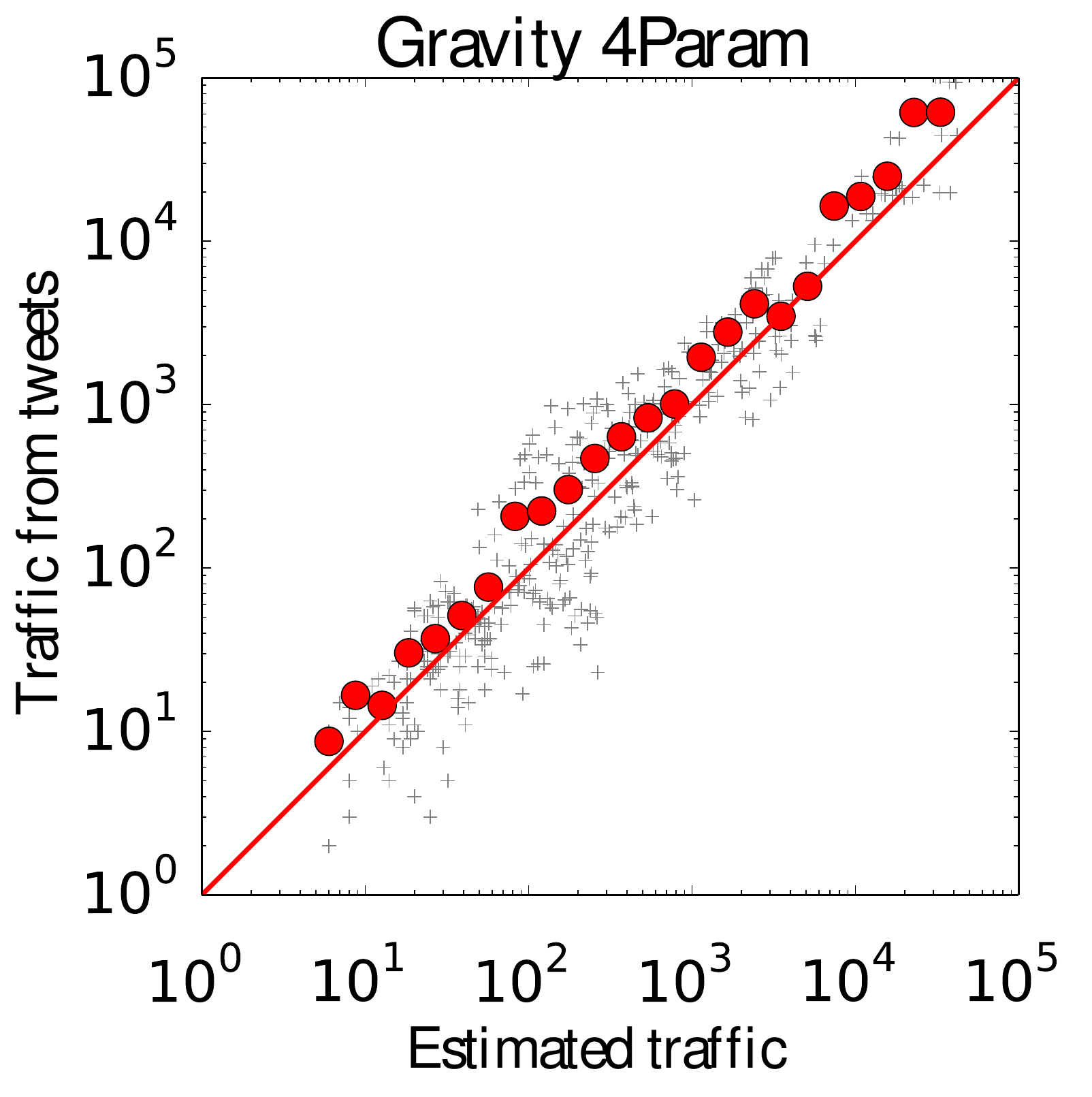}
    \includegraphics[scale = 0.2]{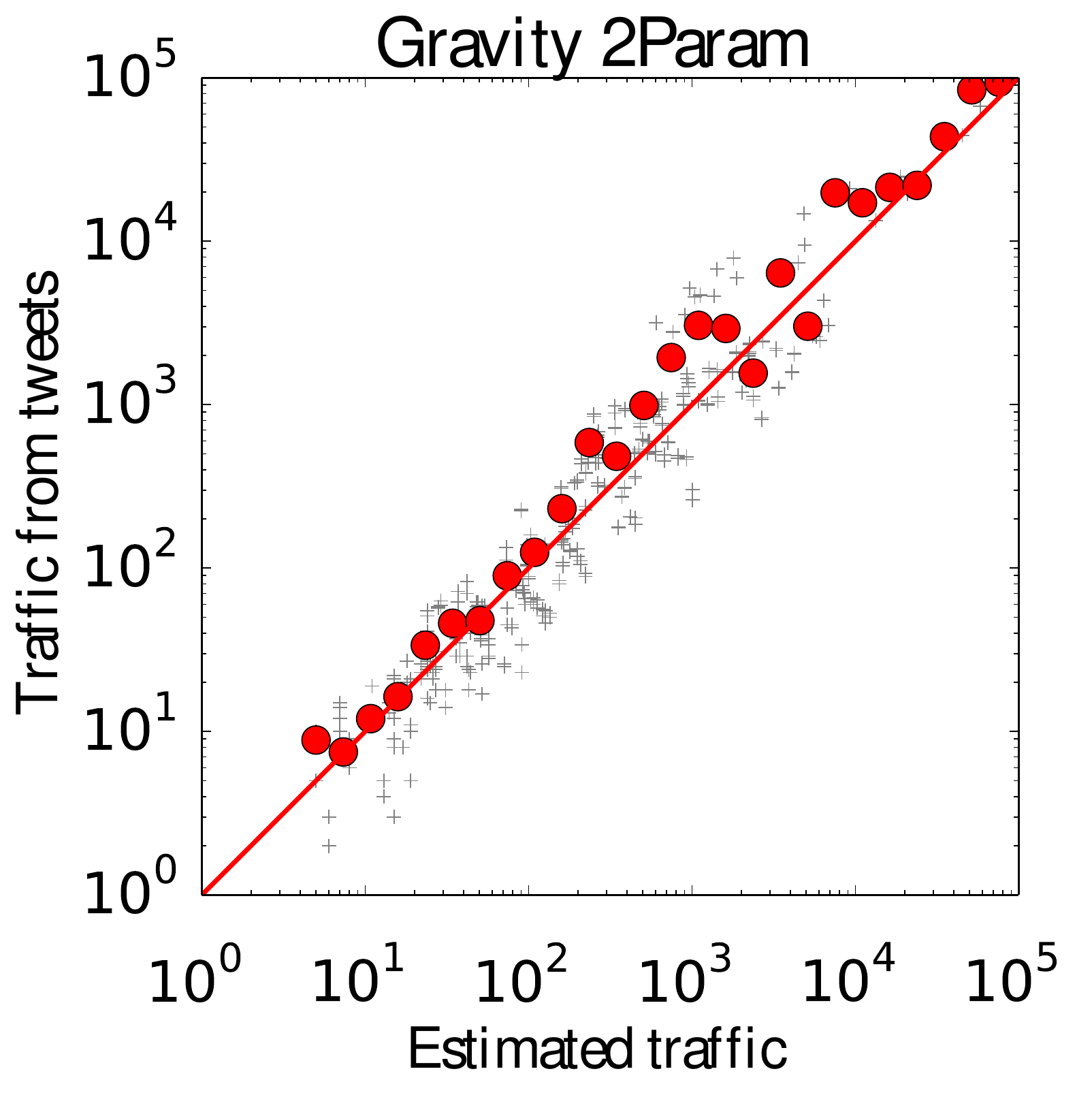}
    \includegraphics[scale = 0.2]{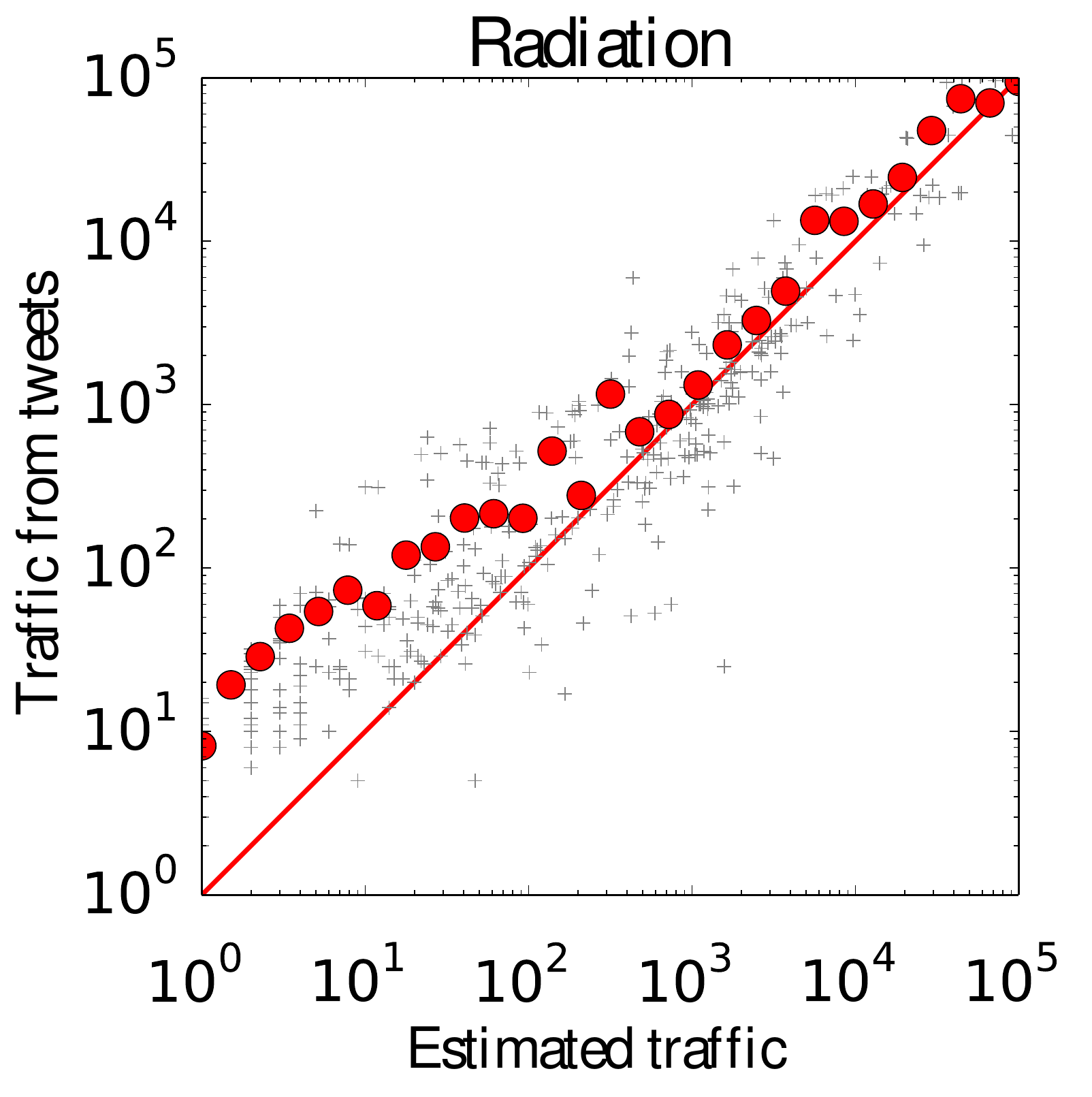}
    \label{fig:modelc}
  }
\caption{Mobility estimation performance of Gravity 4Param, Gravity 2Param and
  Radiation at three scales. Overal Gravity demonstrates
  advantages over Radiation. Radiation's results are
  likely due to the fact that Australia's sparse and uneven population
  distribution deviates from Radiation's implication of a continuous
  dispersion of population from major centers.}
\label{fig:model}
\end{figure}

The performance of the models is reported in Figure
\ref{fig:model}. The $x$-axis is the estimated mobility from the models,
the $y$-axis is the mobility extracted from Tweets, the grey crosses
are the pairs of the estimated and extracted mobility, the red dots are
the averaged values in the bins after logarithmic binning, 
and the red line represents $y=x$. A tighter spread of the 
grey and red markers around the red line
indicates better accuracy, whereas dots falling right of the red
line suggesting overestimation and left suggesting underestimation.

Visually Gravity
models show better performance of capturing the mobility pattern between
areas as well as better robustness to various geographic scales over Radiation. 
For example, For National (Figure
\ref{fig:modela}), it is shown that for Gravity 2Param, with the
exception of a few cases (particularly at larger number scales in the upper
right corner), the grey crosses and red dots tightly surround the red line,
suggesting that in general the estimated mobility is fairly accurate. A
closer examination suggests that the estimation error is roughly
bounded by one decade, with most of the cases being
smaller. Gravity 4Param shows similar distributions
of the grey crosses and the red dots visually, but with slightly greater
dispersions. On the other hand, we could see the grey for Radiation 
scatter loosely around the red line, leading to
a maximal error across almost two decades, with moderate tendency to underestimate.
For State (Figure \ref{fig:modelb}) we note that
Gravity 4Param appears to have the best results, while Gravity 2Param
has a slight tendency to overestimate for larger numbers ($>10^3$).
Radiation yields many overestimations and spans its errors in
three decades. Figure \ref{fig:modelc} indicates that at this scale
both the Gravity models work quite well, while Radiation shows a strong tendency to
underestimate for smaller numbers.

The performance is also
measured quantitatively with two metrics, i.e. the Pearson correlation
coefficient between the estimated mobility and the Twitter mobility, and
the HitRate@50\% (percentage of estimates which have smaller than 50\% relative
errors). Table \ref{tbl:mea} shows these results. For each cell, the
upper number is the Pearson coefficient and the lower the
HitRate@50\%. For each scale and metric, the highest performance is
highlighted. It is evident that, quantitatively, 
Gravity 2Param provides the best performance
overall, which aligns well with the previous conclusions from visual
inspection. 

The results indicate an interesting finding as they contradict the conclusion from \cite{radiation} that Radiation captures human 
mobility better than Gravity. That is, Radiation's advantages are not
universal, and they may not suit countries that have sparsely and unevenly
distributed population, such as Australia or Canada.
Unlike U.S.A. where
a large population spreads relatively evenly across the country,
Australia's population concentrates heavily along its coastline,
creating areas with extremely low population densities between
populated areas. This feature renders Radiation's underlying assumption
that population density decays more smoothly from dense centers
unsuited for Australia.

\vspace{-0.3cm}
\begin{table}[htp]
\centering
\caption{Model performance measured by the Pearson correlation
coefficient (upper) and the HitRate@50\% (lower)}
\begin{tabular}{|c|c|c|c| }
\hline
& Gravity 4Param & Gravity 2Param & Radiation  \\
\hline
National & \begin{tabular}{c}0.877 \\0.330\end{tabular}& \begin{tabular}{c}\textbf{0.912} \\\textbf{0.397}\end{tabular}&\begin{tabular}{c}0.840 \\0.184\end{tabular} \\ 
\hline
State & \begin{tabular}{c}0.893 \\\textbf{0.487}\end{tabular}&\begin{tabular}{c}\textbf{0.896} \\0.397\end{tabular} &\begin{tabular}{c}0.742 \\0.166\end{tabular} \\
\hline
Metropolitan & \begin{tabular}{c}0.948 \\0.53\end{tabular}& \begin{tabular}{c}\textbf{0.963} \\\textbf{0.600}\end{tabular}& \begin{tabular}{c}0.918 \\0.397\end{tabular}\\
\hline
\end{tabular}
\label{tbl:mea}
\end{table}

As we verify the model performance to estimate the mobility extracted
from Tweets, we argue that by replacing $m$ and $n$ with the population
from census, it is feasible to estimate the real-world mobility between
areas in Australia. We will test this
proposal in future work.

\section{Conclusion}
In this paper we explore the feasibility of using geo-tagged Tweets to
estimate population distribution and mobility for Australia. We first
evaluate the correlation between the estimated population to
the census-based population, and then follow with a comparative study modeling
the mobility extracted from Tweets with the gravity model and the
radiation model. To verify the robustness of population estimation and
mobility estimation to different geographic scales, both the
experiments are conducted at three scales that represent the national,
state and metropolitan levels. We report that 1) it is feasible to estimate
population distribution from geo-tagged Tweets, particularly for dense
areas 2) for countries that have uneven population distributions such
as Australia, Gravity model appears a better model to estimate mobility
than Radiation. In future we will further improve the model accuracy
by incorporating census data of higher resolutions, evaluate model
performances with more metrics and at more varieties of distances
scales, 
and use the models
to devise a framework for the prediction of disease spread. 

\bibliographystyle{IEEEtran}
\bibliography{main}

\begin{thebibliography}{10}
\providecommand{\url}[1]{#1}
\csname url@samestyle\endcsname
\providecommand{\newblock}{\relax}
\providecommand{\bibinfo}[2]{#2}
\providecommand{\BIBentrySTDinterwordspacing}{\spaceskip=0pt\relax}
\providecommand{\BIBentryALTinterwordstretchfactor}{4}
\providecommand{\BIBentryALTinterwordspacing}{\spaceskip=\fontdimen2\font plus
\BIBentryALTinterwordstretchfactor\fontdimen3\font minus
  \fontdimen4\font\relax}
\providecommand{\BIBforeignlanguage}[2]{{%
\expandafter\ifx\csname l@#1\endcsname\relax
\typeout{** WARNING: IEEEtran.bst: No hyphenation pattern has been}%
\typeout{** loaded for the language `#1'. Using the pattern for}%
\typeout{** the default language instead.}%
\else
\language=\csname l@#1\endcsname
\fi
#2}}
\providecommand{\BIBdecl}{\relax}
\BIBdecl

\bibitem{census}
\BIBentryALTinterwordspacing
D.~Balcan, V.~Colizza, B.~Gonçalves, H.~Hu, J.~J. Ramasco, and A.~Vespignani,
  ``Multiscale mobility networks and the spatial spreading of infectious
  diseases,'' \emph{Proceedings of the National Academy of Sciences}, vol. 106,
  no.~51, pp. 21\,484--21\,489, 2009. [Online]. Available:
  \url{http://www.pnas.org/content/106/51/21484.abstract}
\BIBentrySTDinterwordspacing

\bibitem{mobile}
\BIBentryALTinterwordspacing
S.~Jiang, G.~A. Fiore, Y.~Yang, J.~Ferreira, Jr., E.~Frazzoli, and M.~C.
  Gonz\'{a}lez, ``A review of urban computing for mobile phone traces: Current
  methods, challenges and opportunities,'' in \emph{Proceedings of the 2Nd ACM
  SIGKDD International Workshop on Urban Computing}, ser. UrbComp '13.\hskip
  1em plus 0.5em minus 0.4em\relax New York, NY, USA: ACM, 2013, pp. 2:1--2:9.
  [Online]. Available: \url{http://doi.acm.org/10.1145/2505821.2505828}
\BIBentrySTDinterwordspacing

\bibitem{cell}
V.~Palchykov, M.~Mitrovi?, H.-H. Jo, J.~Saramäki, and R.~K. Pan, ``Inferring
  human mobility using communication patterns,'' \emph{Scientific Reports},
  vol.~4, no. 6174, 2014.

\bibitem{zhao2011social}
K.~Zhao, J.~Stehl{\'e}, G.~Bianconi, and A.~Barrat, ``Social network dynamics
  of face-to-face interactions,'' \emph{Physical Review E}, vol.~83, no.~5, p.
  056109, 2011.

\bibitem{wifi}
A.~Chaintreau, P.~Hui, J.~Crowcroft, C.~Diot, R.~Gass, and J.~Scott, ``Impact
  of human mobility on opportunistic forwarding algorithms,'' \emph{Mobile
  Computing, IEEE Transactions on}, vol.~6, no.~6, pp. 606--620, June 2007.

\bibitem{meloni2011modeling}
S.~Meloni, N.~Perra, A.~Arenas, S.~G{\'o}mez, Y.~Moreno, and A.~Vespignani,
  ``Modeling human mobility responses to the large-scale spreading of
  infectious diseases,'' \emph{Scientific reports}, vol.~1, 2011.

\bibitem{malaria}
\BIBentryALTinterwordspacing
A.~Wesolowski, N.~Eagle, A.~J. Tatem, D.~L. Smith, A.~M. Noor, R.~W. Snow, and
  C.~O. Buckee, ``Quantifying the impact of human mobility on malaria,''
  \emph{Science}, vol. 338, no. 6104, pp. 267--270, 2012. [Online]. Available:
  \url{http://www.sciencemag.org/content/338/6104/267.abstract}
\BIBentrySTDinterwordspacing

\bibitem{radiation}
F.~Simini, M.~C. González, A.~Maritan, and A.-L. Barabási, ``A universal model
  for mobility and migration patterns,'' \emph{Nature}, vol. 484, pp. 96--100,
  April 2012.

\bibitem{twitter1}
B.~Hawelka, I.~Sitko, E.~Beinat, S.~Sobolevsky, P.~Kazakopoulos, and C.~Ratti,
  ``Geo-located twitter as proxy for global mobility patterns,''
  \emph{Cartography and Geographic Information Science}, vol.~41, no.~3, pp.
  260--271, 2014.

\bibitem{twitter2}
M.~R. Frank, L.~Mitchell, P.~S. Dodds, and C.~M. Danforth, ``Happiness and the
  patterns of life: A study of geolocated tweets,'' \emph{Scientific Reports},
  vol.~3, no. 2625, 2013.

\bibitem{bursts}
A.-L. Barabasi, ``The origin of bursts and heavy tails in human dynamics,''
  \emph{Nature}, vol. 435, p. 207, 2005.

\bibitem{Zipf}
G.~K. Zipf, ``The p1 p2/d hypothesis: On the intercity movement of persons,''
  \emph{American Sociological Review}, vol.~11, no.~6, pp. 677--686, 1946.

\end{thebibliography}

\end{document}